\documentclass[twocolumn,showpacs,preprintnumbers,amsmath,amssymb,aps]{revtex4-1}
\usepackage{graphicx}
\usepackage{epstopdf}           
\usepackage{dcolumn}
\usepackage{bm}
\usepackage[utf8]{inputenc} \usepackage[T1]{fontenc} 
\usepackage{nicefrac}
\usepackage{xfrac}
\usepackage{lmodern}
\usepackage{color}

\definecolor{color}{rgb}{0.11,0.45,0.02}

\usepackage{hyperref}

\usepackage{braket}

\begin{document}

\title{Optical orientation and alignment of excitons in direct and indirect band gap (In,Al)As/AlAs quantum dots with type-I  band  alignment}

\author{J.~Rautert$^1$, T.~S.~Shamirzaev$^{2,3}$, S.~V.~Nekrasov$^4$, D.~R.~Yakovlev$^{1,4}$, P.~Klenovsk\'{y}$^{5-7}$, Yu.~G.~Kusrayev$^4$ and M.~Bayer$^{1,4}$}

\affiliation{
$^1$Experimentelle Physik 2, Technische Universit\"at Dortmund,~44227 Dortmund, Germany \\
$^2$Rzhanov Institute of Semiconductor Physics, Siberian Branch of the Russian Academy of Sciences, 630090 Novosibirsk, Russia \\
$^3$Ural Federal University, 620002 Yekaterinburg, Russia \\
$^4$Ioffe Institute, Russian Academy of Sciences, 194021 St. Petersburg, Russia \\
$^5$Department of Condensed Matter Physics, Faculty of Science, Masaryk University, Kotl\'a\v{r}sk\'a~267/2, 61137~Brno, Czech~Republic\\
$^6$Central European Institute of Technology, Masaryk University, Kamenice 753/5, 62500~Brno, Czech~Republic\\
$^7$Czech Metrology Institute, Okru\v{z}n\'i 31, 63800~Brno, Czech~Republic
}

\begin{abstract}
The spin structure and spin dynamics of excitons in an ensemble of
(In,Al)As/AlAs quantum dots (QDs) with type-I band alignment,
containing both direct and indirect band gap dots, are studied.
Time-resolved and spectral selective techniques are used to
distinguish between the direct and indirect QDs. The exciton fine
structure is studied by means of optical alignment and optical
orientation techniques in magnetic fields applied in the Faraday or
Voigt geometries. A drastic difference in emission polarization is
found for the excitons in the direct QDs involving a $\Gamma$-valley
electron and the excitons in the indirect QDs contributed by an
$X$-valley electron. We show that in the direct QDs the exciton spin
dynamics is controlled by the anisotropic exchange splitting, while
in the indirect QDs it is determined by the hyperfine interaction
with nuclear field fluctuations. The anisotropic exchange splitting
is determined for the direct QD excitons and compared with model
calculations.

\end{abstract}

\maketitle
\section{Introduction}
\label{sec:intro}

Semiconductor quantum dots (QDs) have been studied for many years
due to the broad range of optical applications that they
offer~\cite{Henneberger,Slavcheva}. One of the fundamental
parameters of excitons confined in QDs is their recombination time
that  can  be controlled by  the  type of band
gap~\cite{Shamirzaev96}. Recently, we have demonstrated the
coexistence of both direct and indirect band gap QDs with type-I
band alignment in a (In,Al)As/AlAs semiconductor heterosystem.
This system demonstrates intensive luminescence
 up to room temperature~\cite{ShamirzaevJETP} provided by the strong localization of electrons and holes in (In,Al)As QDs confined by AlAs barriers with a large band gap. Recombination dynamics of indirect excitons is extremely long extending up to several hundreds microseconds at cryogenic temperatures, which allows one to study long living spin
dynamics~\cite{Shamirzaev84}. The difference of direct and indirect
in momentum space excitons was manifested by different optical
techniques: (i) time-resolved photoluminescence (PL) addressing the
recombination dynamics~\cite{Shamirzaev84,ShamirzaevAPL92}, (ii)
optically detected magnetic resonance~\cite{Ivanov97}, and (iii)
spin-flip Raman scattering~\cite{Debus} via the $g$-factors of the
$\Gamma$ and $X$ valley electrons.

In this paper, we study the fine structure of the direct and
indirect excitons in (In,Al)As/AlAs QDs with type-I band alignment.
Time-resolved photoluminescence and luminescence after selective
excitation allow us to distinguish QDs with direct and indirect band
gap in momentum space, which are coexisting also in this QD
ensemble~\cite{ShamirzaevAPL92}. The fine structure is determined
via studies of the polarization of photoluminescence. For the direct
band gap QDs linearly-polarized emission under resonant
linearly-polarized excitation (optical alignment) and the absence of
circularly-polarized emission under resonant circularly-polarized
excitation (optical orientation) are found. However, for the
indirect band gap QDs, unexpectedly, on one hand a negligible
optical alignment and on the other hand pronounced optical
orientation are revealed. The anisotropic electron-hole (e-h)
exchange splitting ($\delta_1$) of the excitons in the direct band
gap QDs is determined. We also demonstrate  that for the indirect
band gap QDs the exciton spin relaxation time ($T_1$) considerably
exceeds the exciton radiative recombination time, which is in the
tens microsecond range.

The paper is organized as follows. In Section~\ref{sec:ExpDetails},
the studied sample and used experimental techniques are described.
In Section~\ref{sec:ES} the emission spectra of an ensemble of
(In,Al)As/AlAs QDs with type-I band alignment, containing QDs with
direct and indirect band gap, are investigated. Using non-resonant
time-resolved and selectively excited time-integrated
photoluminescence spectroscopy we uniquely determine the spectral
features of the excitons confined in the direct and indirect band
gap QDs. Optical alignment and optical orientation of the excitons
is studied as a function of magnetic fields in Sec.~\ref{sec:OA and
OO}. Model calculations of the exciton fine structure are given in
Sec.~\ref{sec:Theory}.

\section{Experimental details}
\label{sec:ExpDetails}

The studied structure (AG 2890) contains 20 layers of undoped
(In,Al)As/AlAs QDs grown by molecular-beam epitaxy on a
(001)-oriented GaAs substrate. The density of the lens-shaped QDs
with an average diameter of $15$~nm and a height of $4$~nm is about
$3 \times 10^{10}$~cm$^{-2}$ in each layer. The QD layers are
separated from each other by $20$-nm-thick AlAs barriers, which
prevent an electronic coupling between QDs in adjacent layers. A
20-nm-thick GaAs cap layer protects the top AlAs barrier against
oxidation. The growth axis is coincided with the crystallographic
direction (001) and defined as $z$-axis. Note, that the band gap
energy of the GaAs substrate is 1.52~eV and that of the AlAs barrier
is 2.30~eV~\cite{Vurgaftman}. Further growth details are given in
Ref.~\cite{Shamirzaev78}.

The sample was mounted either in a liquid helium bath cryostat or
a cold-finger closed-cycle cryostat.  The temperature was varied
from $T = 1.6$ up to 50~K. In the studied QDs the intensity, energy
position and polarization degree of the exciton photoluminescence
were about the same in the temperature range from 1.6 to 10~K. Low
magnetic fields (in the mT range) were generated by an electromagnet and
high magnetic fields (up to 10~T) by a superconducting split-coil
solenoid. The magnetic field was applied either parallel to the
structure growth axis ($\textbf{B} \parallel z$) in the Faraday
geometry or perpendicular to it ($\textbf{B}\bot z$) in the Voigt
geometry. The wave vector of the excitation light was always parallel to
the structure growth axis.

The photoluminescence was excited either non-resonantly with the
photon energy of the laser exceeding considerably the emission
energies in the QD ensemble, or selectively with the laser energy
tuned within the inhomogeneously broadened exciton emission band of QDs.
The non-resonant excitation was provided by the third harmonic of a
Q-switched Nd:YVO$_4$ pulsed laser with photon energy of 3.49~eV, a
pulse duration of 5~ns and a repetition rate of $f=2$~kHz. The
excitation density was kept below 100 nJ/cm$^2$~\cite{Shamirzaev84}.

For selective excitation two lasers were used: (i) an optical
parametric oscillator with tunable photon energy in the spectral
range from 1.5 to 2.0~eV, pulse duration of 1~ps extended by a
multimode optical fiber up to 1~ns and repetition rate of $f=2$~kHz;
(ii) a tunable continuous-wave Ti:sapphire laser. Both techniques
provides similar spectra of QDs, therefore, we do not specify the
used excitation technique in the text.

The photoluminescence was dispersed by an 0.5-m monochromator. For
the time-integrated measurements under non-resonant excitation the
PL was detected by a liquid-nitrogen-cooled charge-coupled-device
(CCD) camera. In order to avoid scattered light from the laser in
experiments with selective excitation we used a gated CCD camera
synchronized with the laser via an external trigger signal. For the
time-resolved PL  at non-resonant excitation, the time delay between
laser pulse and the begin of recording, $t_{\text{delay}}$, was
varied in the range from 1~ns up to 100~$\mu$s. The duration of
recording, i.e. the gate window $t_{\text{gate}}$, was varied from
4~ns to 50~$\mu$s in order to optimize the signal intensity and the
time resolution. The best time resolution of the detection system
was 1~ns.

For measuring the optical alignment and optical orientation effects,
linear or circular polarization of the excitation laser and of the
PL were selected by linear and circular polarizers (Glan-Thompson
prism, as well as quarter-wave and half-wave plates). For the
optical alignment measurement, the linear polarization degree of the
PL ($\rho_\text{l}$) induced under linearly polarized excitation is
measured. The linear polarization degree is defined as
\begin{align}
\rho_\text{l}=\frac{I^\text{0/0}-I^\text{0/90}}{I^\text{0/0} +
I^\text{0/90}},
\end{align}
where $I^{\text{a/b}}$ are the PL intensities with the subscripts
$a/b$ corresponding to the direction of excitation/detection linear
polarization. The direction "0" is parallel to the [110]
crystallographic direction and the direction "90" is parallel to the
$[1\bar 10]$ direction. For the optical orientation measurement, the
circular polarization degree of the PL induced by
circularly-polarized excitation, $\rho_\text{c}$, is measured.
\begin{align}
\rho_\text{c}=\frac{I^{+/+} - I^{+/-}}{I^{+/+} + I^{+/-}}.
\end{align}
Here $I^{\text{a/b}}$ is intensity of the $\sigma^\text{b}$-polarized PL
component measured after $\sigma^\text{a}$-polarized excitation.
The labels $+$ and $-$ correspond to right-hand and left-hand circular
polarization, respectively.

\section{Energy spectra of quantum dots}
\label{sec:ES}

In this paper we study (In,Al)As/AlAs QDs with type-I band
alignment, where the electrons and holes have their minimum energy
 in the (In,Al)As dots and therefore are confined there, see Fig.~\ref{fig1}(b). For all
QD sizes the hole is located in the center of the Brillouin zone. For
large QDs the lowest electron state of the conduction band is in the
$\Gamma$-valley and, therefore, the band gap is direct in momentum
space. In what follows, we will call these QDs the direct QDs.
Contrary to that, in small QDs the quantum confinement shifts the
$\Gamma$ valley of the conduction band in energy above the $X$ valley and the
band gap becomes indirect in momentum space.  We will specify these
QDs as the indirect QDs. It was shown that due to
inhomogeneous broadening an ensemble of (In,Al)As/AlAs QDs is
composed of direct and indirect dots, whose emissions are partly
overlapping~\cite{Shamirzaev78,ShamirzaevAPL92,Debus,Ivanov97}. As we
will show below, the direct and indirect QDs can be well distinguished
from each other by different optical techniques.

\subsection{Time-resolved photoluminescence under non-resonant excitation}
\label{sec:ExpResults_A}

\begin{figure}[!h]
\centering
\includegraphics* [width=8.0cm]{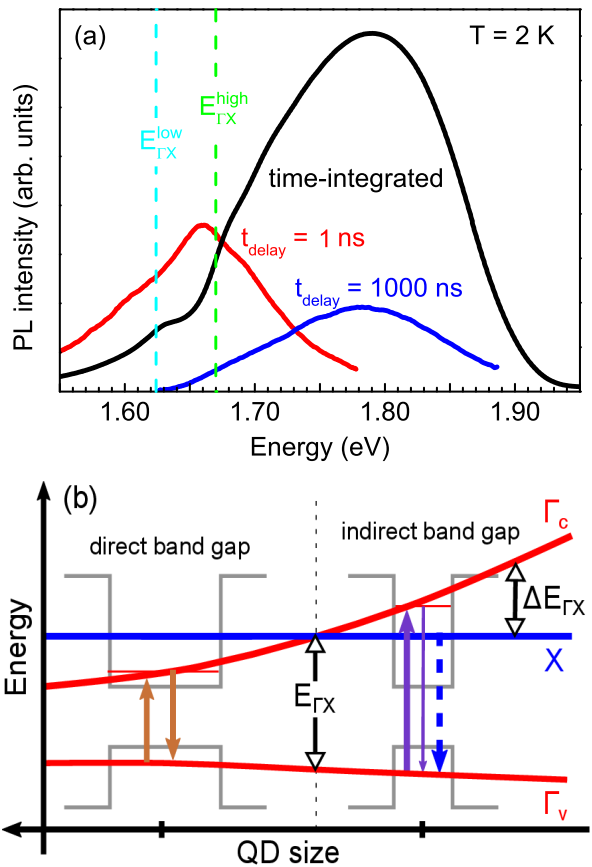}
\caption{(a) Photoluminescence spectra of (In,Al)As/AlAs QDs
measured under non-resonant excitation: time-integrated (black
line), time-resolved for $t_{\text{delay}}=1$~ns and
$t_{\text{gate}}=4$~ns (red), and for $t_{\text{delay}}=1000$~ns and
$t_{\text{gate}}=500$~ns (blue). Vertical dashed lines $E^{\text
{low}}_{\Gamma X}$ (cyan) and $E^{\text {high}}_{\Gamma X}$ (green)
mark ${\Gamma X}$-crossing energies, see
Sec.~\ref{sec:ExpResults_B}. (b) Band structure schematics for two
(In,Al)As/AlAs QDs of different sizes with either the $\Gamma$ or
the $X$ valley as the lowest conduction band state. Vertical arrows
mark the excitation and emission energies. Solid arrows correspond
to direct in momentum space and dashed ones to indirect in momentum
space optical transitions. Dashed vertical line $E_{\Gamma X}$
indicates the QD size with equal energies of the $\Gamma$ and $X$
electron levels. $\Delta E_{\Gamma X}$ is the splitting between the
$\Gamma$ and $X$ electron levels in the conduction band.}
\label{fig1}
\end{figure}

Photoluminescence spectra of an (In,Al)As/AlAs QD ensemble measured
under non-resonant excitation are shown in Fig.~\ref{fig1}(a). The
time-integrated spectrum (black line) has a maximum at 1.80~eV and
extends from 1.55 to 1.95~eV having a full width at half maximum
(FWHM) of 200~meV. The large width of emission band is due to the
dispersion of the QD parameters, since the exciton energy depends on
the QD size, shape and composition~\cite{Shamirzaev78}. The PL band
is contributed by the emission of direct and indirect QDs, which
becomes evident from the time-resolved PL spectra. As they are
measured immediately after the laser pulse application
($t_{\text{delay}}=1$~ns and $t_{\text{gate}}=4$~ns), the PL band
has maximum at 1.66~eV and a FWHM of 120~meV only (red line). For
longer delays ($t_{\text{delay}}=1000$~ns and
$t_{\text{gate}}=500$~ns), the emission shifts to 1.78~eV and
broadens to 190~meV (blue line), rather similar to the
time-integrated PL spectrum.

The strong modification of the time-resolved spectra is provided by
the very different exciton recombination dynamics in direct and
indirect QDs~\cite{Shamirzaev84,ShamirzaevAPL92}. We have
demonstrated recently that after the photoexcitation in the AlAs
barriers electrons and holes are captured in QDs within several
picoseconds, and the capture probability does not depend on the QD
size and composition~\cite{ShamirzaevNT}. Therefore, all QDs in the
ensemble (direct and indirect) become equally populated shortly
after the excitation pulse. The exciton recombination dynamics is
fast for direct QDs, for which emission occurs mainly in the
spectral range of $1.50-1.78$~eV. It is slow for the indirect QDs
emitting in the $1.63-1.95$~eV range. The emission of the direct and
indirect QDs overlaps in the range of $1.63-1.78$~eV. Here, for
further clarification of the relative contributions of the two
different types of QDs to the PL signal, the selective excitation
techniques can be used.

\subsection{Time-integrated photoluminescence under selective excitation}
\label{sec:ExpResults_B}

Selective excitation within an inhomogeneously broadened PL line is
a common technique that allows one to excite only the fraction of
QDs, whose direct exciton transition matches the laser energy
$E_\text{exc}$. As a result, the PL band is transformed into a
spectrum with narrow lines, which is known as fluorescence line
narrowing. For example, in colloidal CdSe quantum dots such a
narrowed PL spectrum consists of the lines of optically-allowed and
optically-forbidden excitons and exciton recombination lines
assisted by LO-phonon emission~\cite{Nirmal,Biadala}. We will show
below that this technique is well suited and informative also for
the studied (In,Al)As/AlAs QDs.

\begin{figure}[!t]
\centering
\includegraphics* [width=9.0cm]{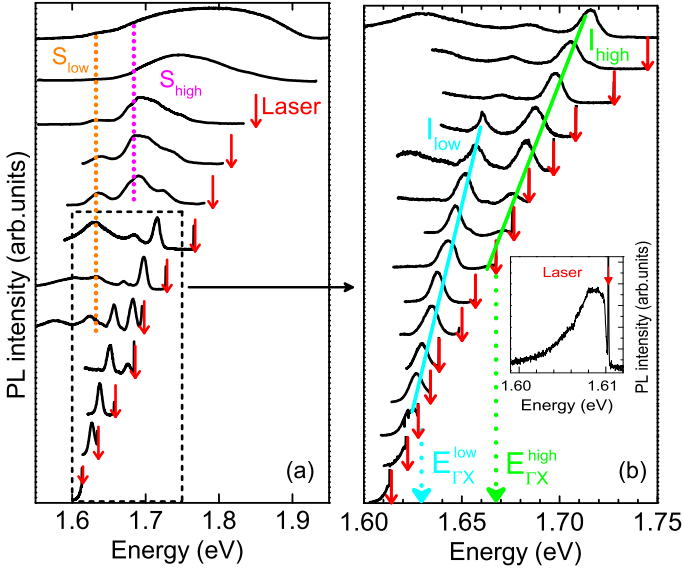}
\caption{(a) Photoluminescence spectra for different excitation
energies indicated by the red arrows. $T=2$~K. Dotted lines indicate
the S$_{\text{low}}$ and S$_{\text{high}}$ PL lines that do not move
with varying laser energy.  (b) Closeup of the PL spectra from panel
(a). Cyan and green lines trace the shifts of the I$_{\text{low}}$
and I$_{\text{high}}$ lines from indirect QDs. Crossing energies of
the PL lines from the direct and indirect QDs are
$E^{\text{low}}_{\Gamma X}=1.63$~eV and $E^{\text{high}}_{\Gamma
X}=1.67$~eV, marked by the vertical dashed arrows. The inset shows
the PL spectrum for $E_\text{exc}=1.61$~eV energy excitation.}
\label{fig2}
\end{figure}

Time-integrated PL spectra measured under
selective excitation tuned in the range of
$E_\text{exc}=1.61-1.95$~eV as well as under non-resonant excitation
at $E_\text{exc}=3.49$~eV (topmost curve) are shown in
Fig.~\ref{fig2}(a). Here, the laser energies are marked by the red arrows.
Scattered light from the laser was excluded by using a gated CCD, see
Sec.~\ref{sec:ExpDetails}.

\begin{figure}[]
\centering
\includegraphics* [width=7.0cm]{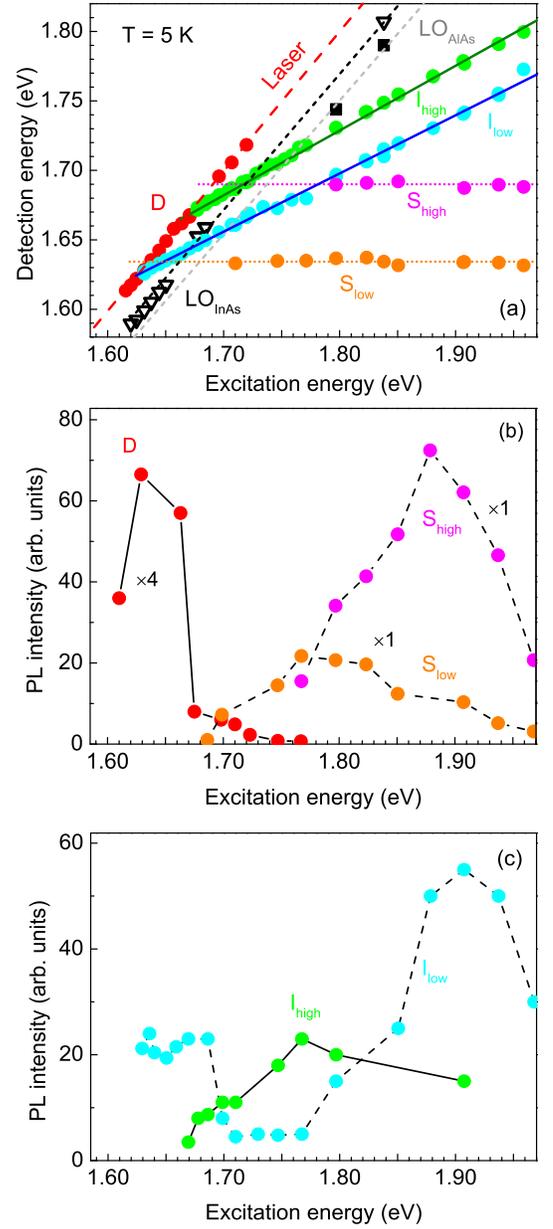}
\caption{Parameters of the PL lines in Fig.~\ref{fig2} plotted as
function of the excitation energy. (a) The shift of the laser is
shown by the red line. Symbols give the energies of the PL lines:
excitons in direct QDs (D-line, red dots); LO-phonon-assisted
emission from direct QDs: LO$_{\mathrm {AlAs}}$ (filled black
squares), LO$_{\mathrm {InAs}}$ (open black triangles). Excitons in
indirect QDs: low energy line (cyan dots) and high energy line
(green dots); S$_{\text{low}}$ (orange dots) and S$_{\text{high}}$
(pink dots) PL lines that do not move with varying laser energy.
Dashed and dotted lines are guides for the eye. Solid lines are fits
using Eqs.~(3) and (4) (see text). (b) Excitation energy
dependencies of the intensity of the D-line (red dots), as well as
the $\text{S}_{\text{low}}$ (orange dots) and
$\text{S}_{\text{high}}$ (magenta dots) PL lines. Lines are guides
for the eye. (c) Excitation energy dependencies of the
I$_{\text{low}}$ (cyan dots) and I$_{\text{high}}$ (green dots) PL
lines. Lines are guides for an eye. } \label{fig3}
\end{figure}

The inset of Fig.~\ref{fig2}(b) shows the PL spectrum for
$E_\text{exc}=1.61$~eV energy excitation. At this energy only the
direct QDs are excited and the recombination energy of the direct
exciton demonstrates a small Stokes shift (about 2 meV) from the
laser photon energy, see the left scheme in Fig.~\ref{fig1}(b).

Also lines with very small intensity shifted to lower energies by 30
and 50~meV appear (not shown). These lines can be
associated with the LO-phonon-assisted recombination of excitons
involving the LO phonons of InAs (30~meV) and AlAs
(50~meV)~\cite{NumeralData}.

When the excitation energy is increased above 1.63~eV, which is on the
low-energy side of the emission of indirect QDs [see
Fig.~\ref{fig1}(a)], additional lines appear in the spectra. As one can
see from Fig.~\ref{fig2}, the number of these lines, their energy
position and intensity depend on the excitation energy. In order to
clarify their properties and origin, we present in Fig.~\ref{fig3}
the peak energy and intensity dependences on the
excitation energy for these lines. These dependencies can be divided into three
groups:

(1) The D-line and its two low-energy satellites related to
LO-photon-assisted recombination with LO$_{\mathrm {InAs}}$ and
LO$_{\mathrm {AlAs}}$ form the first group. These lines follow the
energy shift of the laser. They are associated with exciton
recombination in direct QDs, as illustrated in the left part of
Fig.~\ref{fig1}(b) by the orange arrow, and recombination of excited
$\Gamma$-electron based exciton in indirect QDs (the violet arrow on
the right part of Fig.~\ref{fig1}(b)). With increasing
$E_\text{exc}$ the D-line intensity shows a small increase and then
diminishes by two orders of magnitude up to the excitation energy of
1.75~eV, see Fig.~\ref{fig3}(b). The main reason for that is the
decrease of the direct QD fraction in the ensemble with decreasing
QD size. Also, with increasing $E_\text{exc}$  more photons are
absorbed in the indirect QDs, reducing the excitation efficiency of
the direct QDs. Note, that the indirect and direct QDs have the same
efficiency of light absorption since the absorption is provided by
the direct in momentum space optical transition to the $\Gamma$
valley of the conduction band, as shown by the thick orange and
violet arrows in Fig.~\ref{fig1}(b).

\begin{figure}[!t]
\centering
\includegraphics* [width=6.0cm]{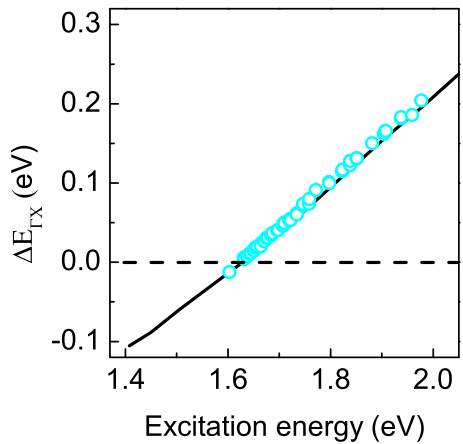}
\caption{Energy difference between the laser photon energy and the
maximum of the I$_{\text{low}}$ line presented as a function of
$E_\text{exc}$ (the symbols). The calculated dependence of $\Delta E_{\Gamma
X}$ in (In,Al)As/AlAs QDs versus excitation energy is shown by the solid line.
The dashed line gives the zero level.} \label{fig4}
\end{figure}

(2) The two lines I$_{\text{low}}$ and I$_{\text{high}}$ shifting to
higher energies with increasing $E_\text{exc}$  with a slope
about twice smaller than that of the D-line and the laser itself form the second group. They are split
off from the D-line at $E^{\text{low}}_{\Gamma X}=1.63$~eV (for
I$_{\text{low}}$) and $E^{\text{high}}_{\Gamma X}=1.67$~eV (for
I$_{\text{high}}$), see the dotted arrows in Fig.~\ref{fig2}(b).
The shifts of these lines with $E_\text{exc}$ can be interpolated by
linear functions that are shown by the lines of corresponding colors in
Fig.~\ref{fig3}:
\begin{align}
E_{\text{low}} ~ \text{[eV]}=0.420\times E_{\text{exc}}+0.942 ,\\
E_{\text{high}} ~ \text{[eV]}=0.461\times E_{\text{exc}}+0.901 .
\end{align}
I$_{\text{low}}$ and I$_{\text{high}}$ lines are provided by exciton
recombination in the indirect QDs. These dots are excited resonantly
via a direct in momentum space optical transition, see the thick
violet arrow in the right part of Fig.~\ref{fig1}(b). Their
emission, shown by the dashed blue arrow, is shifted from the
excitation energy by the energy difference between the $\Gamma$ and
the $X$ valleys in the conduction band, $\Delta E_{\Gamma X}$, which
increases with increasing $E_\text{exc}$. This dependence is shown
for the I$_{\text{low}}$ line by the symbols in Fig.~\ref{fig4} and
is compared with the results of model calculations (the solid line)
using the simple-band effective-mass approach described in
Ref.~\cite{Shamirzaev78}, which takes into account
strain, deformation potentials and nonparabolicity of the electron
dispersion~\cite{Kane}. In this calculation we also take into account dependencies of the $\Gamma$ and $X$ electron levels on the QD size and
shape (obtained from microscopy data), alloy composition and its
gradient across a QD (obtained from growth condition using nomogram
from Ref.~\cite{Shamirzaev78}). One sees excellent agreement
between the experimental data and the calculations, which is
remarkable as we do not use any fitting parameters. The intensities
of I$_{\text{low}}$ and I$_{\text{high}}$ lines have non-monotonic
dependencies on the excitation energy, as shown in
Fig.~\ref{fig3}(c).

It is surprising to observe two emission lines related to the indirect QDs.
We explain that by a bimodal distribution in the QD ensemble. Possibly
the studied sample contains two subensembles of QDs
with different shape and/or composition. Each of them is
characterized by its own $E_{\Gamma X}$. The reason for such a bimodal
distribution is still not clarified, most probably it is determined
by the sample growth conditions.

\begin{figure}[t]
\centering
\includegraphics* [width=9.2cm]{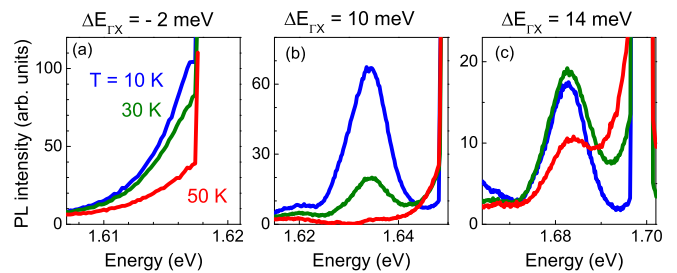}
\caption{Evolution of PL spectra with increasing temperature for (a)
D-line at $E_\text{exc}=1.619$~eV, (b) I$_{\text{low}}$ line at
$E_\text{exc}=1.645$~eV, and (c) I$_{\text{high}}$ line at
$E_\text{exc}=1.696$~eV.} \label{fig5}
\end{figure}

Let us discuss the role of the direct emission from the indirect
QDs. After resonant excitation through the direct optical
transition, the direct exciton contributes to the emission at the
laser photon energy. However, at liquid helium temperatures this
state is quickly depopulated by electron relaxation to the $X$
valley. We showed that the time of this relaxation is much shorter
than the radiative recombination time of the direct
exciton~\cite{ShamirzaevNT}. However, at elevated temperatures, when
the $\Gamma$ valley becomes thermally populated, observation of the
direct emission from the indirect QDs is possible. In
Figs.~\ref{fig5}(b) and \ref{fig5}(c) one sees a redistribution of
the emission intensity from the indirect to the direct recombination
channel when a temperature is increased. The latter forms the
low-energy wing close to the excitation energy. For comparison, we
show in Fig.~\ref{fig5}(a) the temperature evolution of the D-line
in the direct QDs. Here only a decrease of the PL intensity is
observed with raising temperature which is explained by the opening
of nonradiative decay channels.

(3) The two lines $\text{S}_{\text{low}}$ and
$\text{S}_{\text{high}}$, which energies do not depend on the
excitation energy [Figs.~\ref{fig2}(a) and \ref{fig3}(a)], are in the third group. They are relatively broad lines. The $\text{S}_{\text{low}}$ line has the peak energy at 1.635~eV and appears in the spectra for excitation energies
exceeding 1.70~eV. The $\text{S}_{\text{high}}$ line has the peak
energy at 1.690~eV and appears for $E_\text{exc}>1.80$~eV. The
dependencies of their intensities on the excitation energy are shown
in Fig.~\ref{fig3}(b).

These lines show up in the time-integrated spectrum (Fig.~\ref{fig1}(a)) in the spectral range corresponding to
the ${\Gamma\text{X}}$ transition for the $\text{I}_{\text{low}}$ and
$\text{I}_{\text{high}}$ lines. The origin of the
$\text{S}_{\text{low}}$ and $\text{S}_{\text{high}}$ lines is still
unknown and is beyond the scope of this study.

\section{Optical alignment and optical orientation}
\label{sec:OA and OO}

We have shown in Sec.~\ref{sec:ES} that the D,
$\text{I}_{\text{low}}$ and $\text{I}_{\text{high}}$ PL lines can be
identified with excitons in direct and indirect QDs. This opens a
way to study and compare the fine structures of these excitons using
selective excitation together with optical alignment and optical
orientation.

Note that for small values of $\Delta E_{\Gamma X}$ an
admixing of $\Gamma$ and X-electron states takes place~\cite{Debus}.
However, the effect of this admixing on the exciton fine structure  is not
in the focus of this paper.

\subsection {Direct and indirect QDs in zero magnetic field}
\label{sec:ExpResults_C}

\begin{figure}[ht]
\includegraphics* [width=8.0cm]{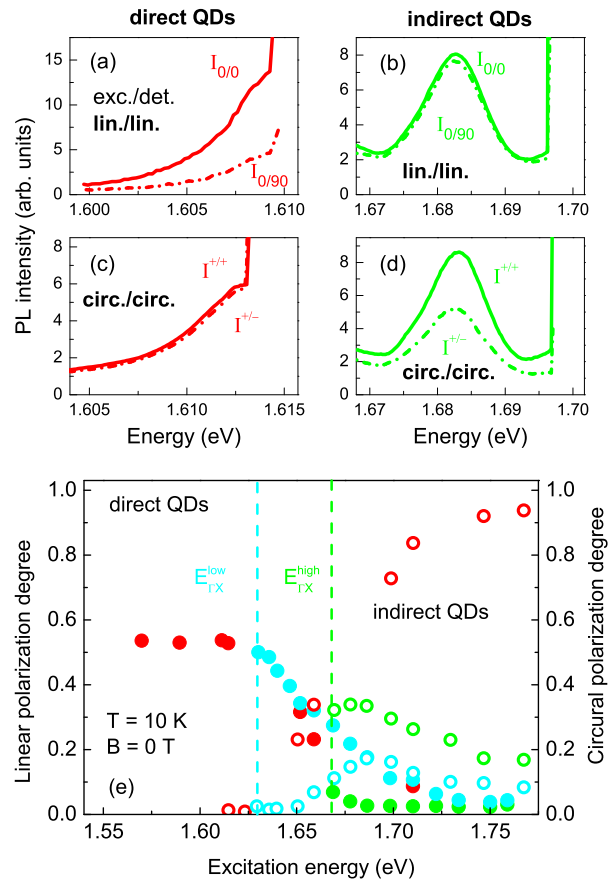}
\caption{Linear polarized PL intensities of direct (a) and indirect
(b) excitons  under linear polarized excitation. Circular polarized
PL intensities of direct (c) and indirect (d) excitons under
circular polarized excitation. $E_{\text{exc}}= 1.61$~eV for the
direct QDs and $E_{\text{exc}}=1.70$~eV for the indirect QDs.
$T=10$~K. (e) Linear (filled circles) and circular (open circles)
polarization degree of the emission for optical alignment and
orientation, respectively, measured for the direct (D-line, red
symbols) and indirect excitons (lines I$_{\text{low}}$ and
I$_{\text{high}}$, cyan and green symbols, respectively) as a
function of excitation energy. Vertical dashed lines mark the
intersection range for PL from the direct and indirect QDs. }
\label{fig6}
\end{figure}

Let us start with experiments performed at zero magnetic field. We
choose two particular excitation energies to address either direct
($E_{\text{exc}}=1.61$~eV) or indirect ($E_{\text{exc}}=1.70$~eV)
QDs and measure the respective optical alignment and optical
orientation. Results for polarized PL spectra are shown in the four
upper panels of Fig.~\ref{fig6}. One can see, that the properties of
the direct and indirect QDs are very different with respect to
optical alignment and orientation. In the direct QDs 53\% of
linearly polarization of PL is detected in the case of linear
polarized excitation, while the circular polarization in the case of
circular polarized excitation is negligible, compare
Figs.~\ref{fig6}(a) and \ref{fig6}(c). Contrary to that, in the
indirect QDs the optical alignment does not exceed 2\%, but the
optical orientation reaches 33\%, Figs.~\ref{fig6}(b) and
\ref{fig6}(d).

These results can be explained by considering the exciton fine
structure. It is known that for axial symmetrical QDs the exciton
states are four-fold degenerate in case of (hypothetical) absence of
the electron-hole exchange interaction. A nonzero isotropic exchange
interaction splits the exciton state, which is characterized by the
total angular momentum projections $J_z = \pm 1, \pm 2$, into the
doubly-degenerate bright and doubly-degenerate dark exciton states
with $J_z = \pm 1$ and $J_z = \pm2$, respectively. The breaking of
the axial symmetry in real QDs lifts the degeneracy of the bright
exciton states and mixes them so that the following states emerge:
$\ket{\text{X}}=\frac{1}{\sqrt{2}}(\ket{+1}+\ket{-1})$ and
$\ket{\text{Y}}=\frac{1}{i\sqrt{2}}(\ket{+1}-\ket{-1})$~\cite{Bayer}.

A circular polarized photon excites the superposition states
$\ket{\text{X}}$ and $\ket{\text{Y}}$, which coherence is rapidly
lost with time~\cite{Paillard}. The decoherence time equals
$\hbar/\delta_1$, where $\delta_1$ is the splitting of the
$\ket{\text{X}}$ and $\ket{\text{Y}}$ doublet by the anisotropic
part of the electron-hole exchange interaction. If the decoherence
time is much shorter then the exciton recombination time,
$\tau_\text{R}$, the optical orientation of the excitons is
destroyed. In contrast to circular polarized light, linear polarized
photons excite the pure states $\ket{\text{X}}$ and $\ket{\text{Y}}$
of the direct exciton, so that the linear polarization degree of the
emission (optical alignment) is controlled by the ratio of the
exciton spin decoherence time to $\tau_\text{R}$. The high value
more than 50\% of optical alignment shown for the direct QDs in
Fig.~\ref{fig6}(a), leads us to the conclusion that the spin
decoherence time exceeds  the recombination one, which is typical
for the direct band gap QDs~\cite{Paillard}.

In indirect QDs the exchange interaction is weak because the wave
functions of the $X$-electron and the $\Gamma$-hole overlap weakly
in momentum space~\cite{Pikus,Bir74}. Thus, the mixing of the
electron spin states is decreased and the pure exciton spin states
$\ket{\pm 1}$ are formed. Therefore, in the indirect QDs with small
or absent anisotropic exchange splitting $\delta_1$, a larger
circular polarization degree can be realized. This is in line with
our experimental data presented in Fig.~\ref{fig6}(d).

We have performed a detailed study of the spectral dependence of the
optical alignment and optical orientation in the range of excitation
energies  from 1.57 to 1.77~eV. The results for the direct (D-line)
and indirect (I$_{\text{low}}$ and I$_{\text{high}}$ lines) excitons
are summarized in Fig.~\ref{fig6}(e).

The optical alignment of the D-line (direct exciton, marked by red
solid circles) has a plateau of $\rho_\text{l}=53\%$ in the range of
$E_{\text{exc}}=1.57-1.61$~eV. Near the crossing energy
$E^{\text{low}}_{\Gamma X}$ the $\rho_\text{l}$ starts to decrease.

In indirect QDs the optical alignment is also strongly dependent on
the excitation energy. The low-energy indirect exciton
(I$_{\text{low}}$ line, solid cyan circles) has $\rho_\text{l}=50\%$
at the crossing energy $E^{\text{low}}_{\Gamma X}$, which is the
same as for the D-line in the direct QDs, compare the solid cyan and
red circles in Fig.~\ref{fig6}(e). For larger $E_{\text{exc}}$ the
linear polarization degree decreases monotonically down to a
constant level of $\rho_\text{l}=2\%$ for excitation energies
exceeding 1.75~eV. The second indirect optical transition
(I$_{\text{high}}$ line, solid green circles) has
$\rho_\text{l}=6\%$ at its appearance energy
$E^{\text{high}}_{\Gamma X}$,  which quickly decreases down to 2\%
for $E_{\text{exc}}> 1.69$~eV.


Let us turn to the spectral dependence of the optical orientation
shown by the open circles in Fig.~\ref{fig6}(e). For the D-line
(red) it is zero for $E_{\text{exc}} \leq 1.62$~eV. Its increase
starts at $E^{\text{low}}_{\Gamma X}$ with the transition from
direct to indirect QDs and $\rho_\text{c}$ reaches 95\% at
$E_{\text{exc}}>1.75$~eV. Note, that the increase in $\rho_\text{c}$
is correlated with the decrease of the D-line intensity,
Fig.~\ref{fig3}(b). Both dependences arise from the decreasing
lifetime of the direct exciton in indirect QDs, as discussed in
Sec.~\ref{sec:ExpResults_B}, which is controlled by its fast
scattering into the indirect exciton state. The optical orientation
degree $\rho_\text{c}$ is determined by the ratio of the exciton
spin relaxation time to its lifetime. In direct QDs the exciton
lifetime, being controlled by the radiative recombination, is longer
than the decoherence time $\tau_\text{d}$, which is inversely
proportional to $\hbar/\delta_1$, so that circular polarization
during the exciton lifetime disappears and $\rho_\text{c} \approx
0$. However, in indirect QDs the lifetime of the direct excitons
becomes very short, which provides a revival of the large optical
orientation observed even in QDs with considerable anisotropic
exchange splitting $\delta_1$ of the direct exciton, which is not
the ground state in the indirect dots.

The optical orientation of the indirect excitons in the indirect QDs
is shown in Fig.~\ref{fig6}(e) by the cyan and green open circles
for the I$_{\text{low}}$ and I$_{\text{high}}$ lines, respectively.
The circular polarization degree of the I$_{\text{low}}$ increases
with growing $E_{\text{exc}}$ and reaches $\rho_\text{c} = 17\%$ at
1.68~eV. It shows a decrease down to 9\% with further rise of the
excitation energy. The I$_{\text{high}}$ line reaches its maximum
optical orientation of 33\% at 1.675~eV, i.e. shortly above the
appearance of this line at $E^{\text{high}}_{\Gamma X}$. Then it
decreases down to 17\% for the excitation energy reaching 1.77~eV.
This decrease can be explained by an enhanced depolarization of the
excitons caused by their scattering from the direct to the indirect
state, i.e., by the electron scattering from the $\Gamma$ to $X$
valley before exciton recombination. The energy dissipated by this
scattering equals to $\Delta E_{\Gamma X}$. It increases for smaller
QDs requiring, e.g., more acoustic phonons, which in turn enhances
the perturbation of the exciton spins.

\subsection{Effect of the magnetic field on optical alignment and orientation in direct QDs}

\begin{figure}[!t]
\centering
\includegraphics* [width=8.0cm]{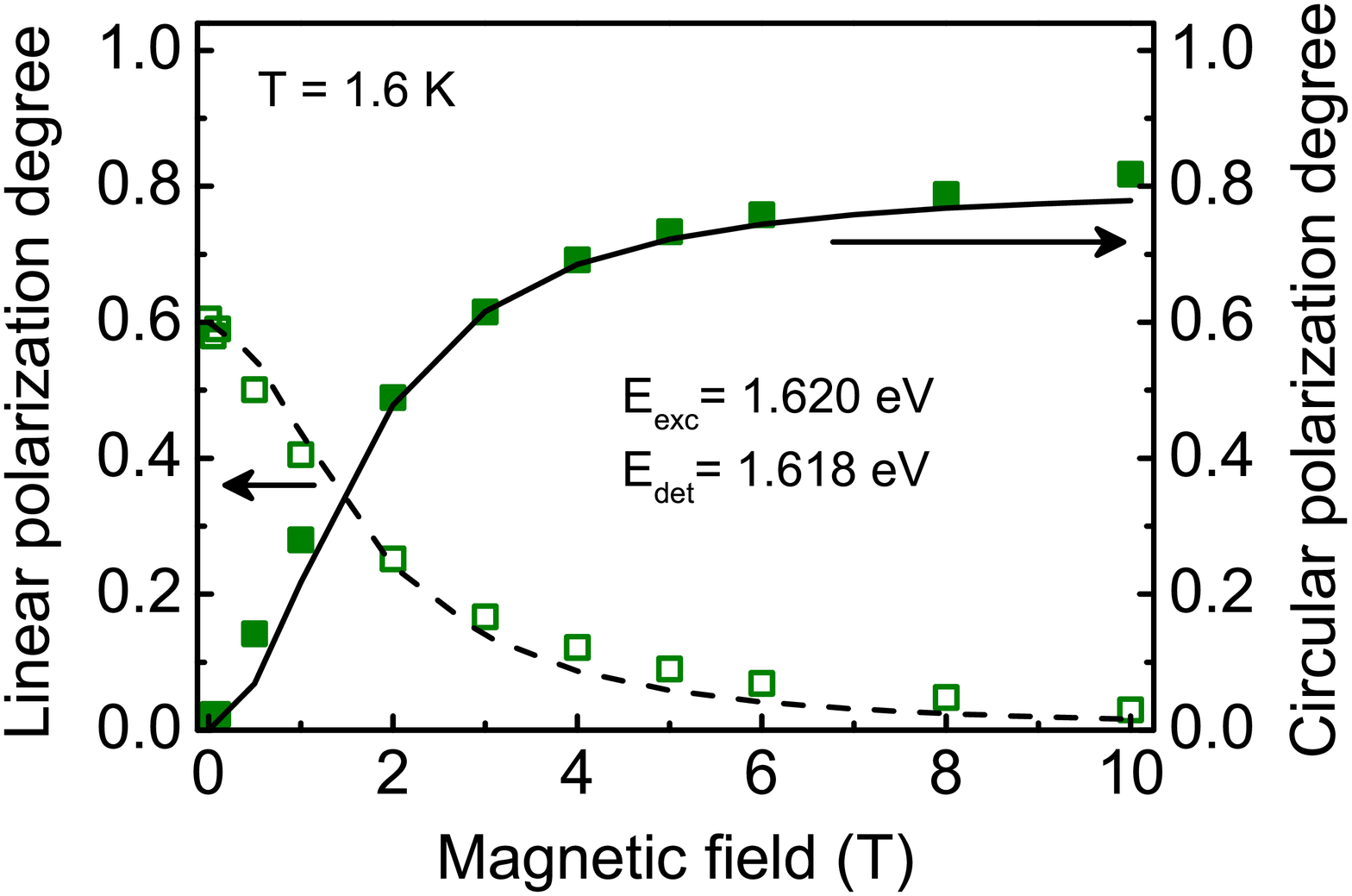}
\caption{Degree of linear (open circles) and circular (filled
circles) polarization of the direct exciton emission (D-line) as a
function of longitudinal magnetic field (Faraday geometry). Lines
show fits to experimental data with Eqs.~\eqref{eq:a1} and
\eqref{eq:a2} using the single fitting parameter $\delta_1=260\pm
10$~$\mu$eV. } \label{fig7}
\end{figure}

The application of magnetic field is a powerful tool for the
investigation of the exciton fine structure. The results on optical
alignment and optical orientation in strong magnetic fields up to
10~T are shown in Fig.~\ref{fig7}. Here the direct QDs were excited
at $E_{\text{exc}}=1.620$~eV. The magnetic field was applied in
Faraday geometry ($\textbf{B} \parallel z$). One sees that the
optical alignment is suppressed by the magnetic field, as the linear
polarization degree decreases from $\rho_\text{l}=59\%$ at zero
field down to 2\% for $B=10$~T. On the other hand, the optical
orientation, which is absent at zero field, increases with
increasing field and saturates at $\rho_\text{c}=80\%$.

These experimental data are in a good agreement with the theory. It
was demonstrated in Ref.~\cite{Dzhioev} that the longitudinal
magnetic field suppresses the exciton spin alignment and restores of
the exciton spin orientation. These effects occur when the Zeeman
splitting between the $\ket{+1}$ and $\ket{-1}$ exciton states
exceeds the energy of the anisotropic exchange interaction
$\delta_1$, which mixes these states. The magnetic field dependences
of the linear and circular polarization degrees can be described by
the following equations:
\begin{align}
\label{eq:a1}\rho_\text{l}(B) = \rho^{0}_\text{l}
\frac{\delta_1^{2}}{\delta_1^{2}+(\mu_{\text{B}} g_\text{ex} B)^2} , \\
\label{eq:a2}\rho_\text{c}(B) = \rho^{\text{max}}_\text{c} \frac{(\mu_{\text{B}} g_\text{ex}
B)^2}{\delta_1^{2}+(\mu_{\text{B}} g_\text{ex} B)^2} ,
\end{align}
where $\rho^{0}_\text{l}$ is the linear polarization degree at zero
magnetic field, and $\rho^{\text{max}}_\text{c}$ is the maximal
circular polarization degree achieved in strong fields,
$\mu_{\text{B}}$ is the Bohr magneton, and $g_\text{ex}$ is the
longitudinal exciton $g$-factor. The latter is composed of the
$\Gamma$-electron ($g_\text{e}$) and heavy hole ($g_\text{hh}$)
$g$-factors: $g_\text{ex}=g_\text{hh}-g_\text{e}$. Both dependencies
of $\rho_\text{l}(B)$ and $\rho_\text{c}(B)$  given by
Eqs.~\eqref{eq:a1} and \eqref{eq:a2} have a Lorentz shape and the
same FWHM of $B_{1/2}= {2{\delta_1}/ \mu_{\text{B}} g_\text{ex}}$.
We fit the experimental data for the direct QDs with
Eqs.~\eqref{eq:a1} and \eqref{eq:a2}, see the lines in
Fig.~\ref{fig7}. The exciton $g$-factor $g_\text{ex}=2.63$, as
evaluated from the known values for $g_\text{hh}=2.43$~\cite{Debus}
and $g_\text{e}=-0.2$~\cite{Yugova}. The only free parameter in the
fit was $\delta_1$, which allows us to determined it to
$\delta_1=260\pm 10$~$\mu$eV, corresponding to $B_{1/2}=
3.1\pm0.15$~T. We theoretically confirm the value of the anisotropic
exchange energy $\delta_1$ and show that it agrees with calculations
in Sec.~\ref{sec:Theory}.
%

\subsection{Effect of the magnetic field on the optical orientation in indirect QDs}

\begin{figure}[!t]
\centering
\includegraphics* [width=8.0cm]{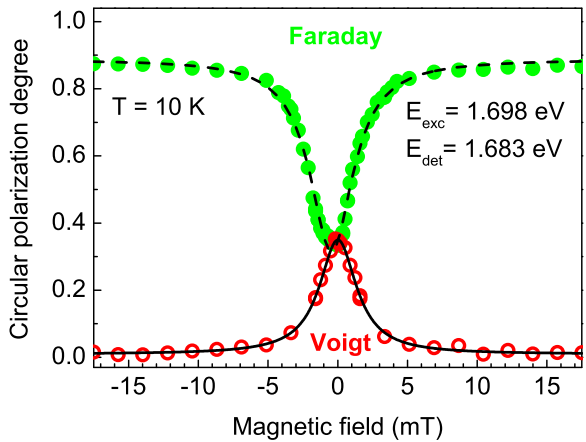}
\caption{Circular polarization degree of the I$_{\text{high}}$  line
as function of magnetic field in Faraday (green filled circles) and
Voigt (red open circles) geometry that are referred in the
text as polarization recovery curve and Hanle curve, respectively.
Solid and dashed lines are fits by Lorentz curves with FWHMs of 3.2
and 3.8 mT, respectively.} \label{fig8}
\end{figure}

Here we focus on the optical orientation of the indirect excitons
in the indirect QDs. We choose $E_{\text{exc}}=1.698$~eV and focus on the
I$_{\text{high}}$ line with maximum at 1.683~eV. For this line the
optical alignment is absent and the optical orientation degree at
zero magnetic field is $\rho_\text{c}=33\%$. $\rho_\text{c}(B)$
demonstrates strong changes already in weak magnetic fields of a few mT
and the changes depend on the field orientation, see
Fig.~\ref{fig8}.

For the Voigt orientation of the magnetic field ($\textbf{B}\bot z$)
the Hanle effect on the optical orientation is observed. Namely, the
circular polarization degree is reduced down to zero with growing field . This decrease is well described by a Lorentz curve with a FWHM
of 3.2~mT. When the magnetic field is applied in Faraday geometry
($\textbf{B} \parallel z$), an increase of the circular polarization
degree up to 89\% takes place. This dependence can be also described
by a Lorentz curve with FWHM of 3.8~mT, which in fact coincides with
the width of the Hanle curve. Note, that in the Faraday geometry the
minimum of $\rho_\text{c}(B)$ is slightly shifted by $-0.45$~mT from
the zero field value. We explain that by dynamic nuclear
polarization~\cite{OO_book}.

The experimentally observed specifics of the optical orientation in
indirect QDs presented in Fig.~\ref{fig8} are characteristic for the
electron spin dynamics in QDs controlled by frozen nuclear spin
fluctuations~\cite{Merkulov,Smirnov2018}. It is expected that in
this case the widths of the Hanle curve and polarization recovery
curve coincide. Also, the polarization recovery in Faraday geometry
increases the polarization degree by a factor of 3 being in line
with model predictions. Note, that the model considerations of
Refs.~\onlinecite{Merkulov,Smirnov2018} are formulated for an
electron confined in a QD, but they can be readily used for our case
of an electron in an indirect exciton, as for the $X$-electron the
exchange interaction with $\Gamma$ hole is very weak and can be
neglected~\cite{Pikus,Bir74}.

The finite number of nuclear spins in a QD results in nuclear spin
fluctuations. Their spin dynamics is very slow compared to the
electron spin dynamics and, therefore, for the electron these
fluctuations can be considered as frozen effective magnetic fields
$\mathbf{B}_N$. The electron spin undergoes Larmor precession around
$\mathbf{B}_N$ with a frequency $\mathbf{\Omega}_N = \mu_\text{B}
g_\text{e} \mathbf{B}_N/ \hbar$. If the condition $\Omega_N
\tau_\text{c} >>1 $ is valid, where $\tau_\text{c}$ is correlation
time in the fluctuating magnetic fields, then the photogenerated
100\%  spin-oriented electrons lose 2/3 of their
spin polarization as $\mathbf{B}_N$ has no preferable spin
orientation and its direction vary from dots to dots. The rest 1/3
of the electron polarization is stabilized via interaction with the
nuclear spins pointing along the orientation direction
that results in a PL polarization degree of about 33\%. In
experiment, the transverse magnetic field in Voigt
geometry decreases this spin polarization. In this
case it is in competition with the stabilization action of
$\mathbf{B}_N$ longitudinal component. As a result,
$B_{1/2}$ of the Hanle curve should be equal to $B_N$. The
longitudinal magnetic field in Faraday geometry
stabilizes the 2/3 of the electron polarization that would be
destructed by $\mathbf{B}_N$ transverse component at
zero field. In this case the field is again in competition with the
nuclear fluctuations and $B_{1/2}=B_N$. It is worth to note, that
the spin polarization recovers to the very high value of 89\%, which
means that in magnetic fields exceeding 6~mT (Faraday geometry) the
exciton spin relaxation time is considerably longer than the
lifetime of the indirect exciton (up to 30~$\mu$s, that correspond
to maximum of the radiative exciton recombination times distribution
in the QD ensemble found by technique proposed in
Ref.~\cite{Shamirzaev84}).

\begin{figure}[!t]
\centering
\includegraphics* [width=7.5 cm]{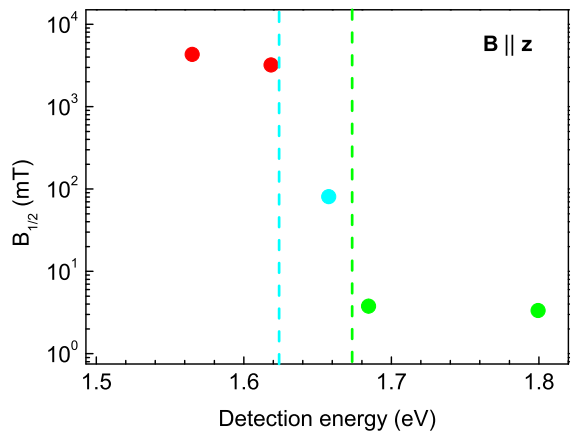}
\caption{ Width $B_{1/2}$ of the longitudinal magnetic field
dependencies of the circular polarization degree (Faraday geometry)
measured in the direct QDs (D-line, red dot), and the indirect QDs
(I$_{\text{low}}$ and I$_{\text{high}}$ lines, cyan and green dots,
respectively). Vertical dashed lines mark the intersection range of
PL from the direct and indirect QDs.} \label{B1/2_Edet}
\end{figure}

One also notes by comparing the data from Figs.~\ref{fig7} and
\ref{fig8}, that the characteristic $B_{1/2}$ values differ by about
three orders of magnitude for the excitons in the direct and
indirect QDs. In Fig.~\ref{B1/2_Edet} we plot the dependence of
$B_{1/2}$ on the excitation energy. One can see that similar to
other properties, like spectral shifts and polarizations, there are
distinct spectral ranges for the direct and indirect QDs.

\section{Calculation of anisotropic exchange energy}
\label{sec:Theory}

\begin{figure}[!h]
\includegraphics* [width=8.0cm]{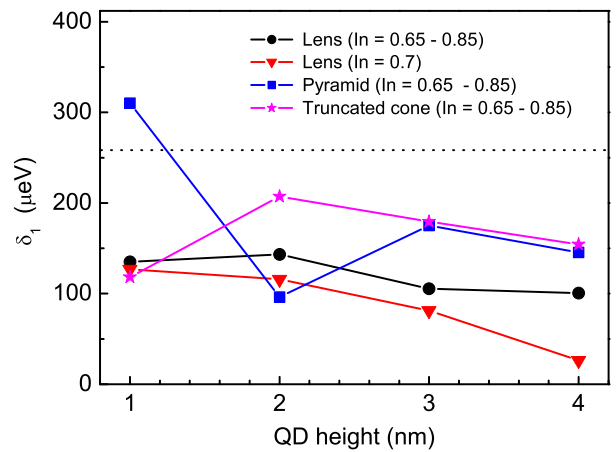}
\caption{Calculation of $\delta_1$ for the direct exciton in
(In,Al)GaAs/AlAs QD as a function of QD height for lens, pyramid,
and truncated cone shaped QDs with an In content either of 0.7 set
constant in the whole QD volume or linearly increasing along the dot
vertical symmetry axis from 0.65 at QD base to 0.85 at QD apex. In
all cases the QD base diameter was fixed to 15~nm. The data include
EX$_0$+EX$_1$+EX$_2$ contributions in the exchange interaction, see
text for details. The dotted horizontal line marks the value of
$\delta_1=260\,\,\mu\rm{eV}$ obtained from fitting of the
experimental data.} \label{Theory_FSS}
\end{figure}

The exciton states in the (In,Al)As/AlAs direct quantum dots were
calculated by first computing the single particle states using the
envelope function approximation based on the 8-band ${\bf k \cdot
p}$ method for both electrons and holes, using the nextnano++
simulation suite~\cite{nextnano}. The obtained wavefunctions were
then utilized as basis states for the configuration interaction (CI)
method that takes into account the corrections due to the direct and
exchange Coulomb interaction. The whole method is detailed in
Ref.~\cite{Klenovsky2017}. We consider here the exchange interaction
up to second order of the multipole
expansion~\cite{TAKAGAHARA1993,Krapek2015}. Following
Ref.~\cite{Krapek2015} we mark the terms of that expansion as
follows: monopole-monopole (EX$_0$), monopole-dipole (EX$_1$), and
dipole-dipole (EX$_2$). The theoretical results of $\delta_1$ for
different sizes and shapes  of QD with mean In content of 0.7 are
shown in Fig.~\ref{Theory_FSS}.

Clearly, Fig.~\ref{Theory_FSS} shows that our calculations to a good
approximation reproduce the experimental value of $\delta_1$ deduced
from fitting of the experimental data by
Eqs.~(\ref{eq:a1})~and~(\ref{eq:a2}). Moreover, our calculations
confirm the assumption that $\delta_1$ depends on QD asymmetry (note
e.g., the larger values of $\delta_1$ for an alloy gradient compared
to the constant In value in Fig.~\ref{Theory_FSS}).\\

Finally we note that the magnitude of the anisotropic
exchange interaction in indirect QDs is at least smaller than
${\delta_{\text{max}}}$= ${\sfrac{1}{2} B_\text{N} \mu_\text{B} g_{\text{ex}} }$, otherwise, the optical
orientation in these QDs will be destroyed by the mixing of exciton states. Taking into account that ${B_{1/2}}$ of the Hanle curve should be equal to
${B_\text{N}}$ and $g_{\text{ex}}$ for indirect exciton in (In,Al)As/AlAs QDs
equals to 0.43~\cite{Debus} we can estimate the upper limit of the
exchange splitting for an exciton in indirect QDs as $\delta_{\text{max}} = 4\times 10^{-8}$~eV.

\section{Conclusions}
\label{sec:conclusions}

The energy spectra and exciton fine structure in an (In,Al)As/AlAs
QD ensemble with type-I band alignment with direct and indirect band
gap have been investigated. Exciton optical alignment (orientation)
was observed for direct (indirect) band gap QDs. The anisotropic e-h
exchange energy $\delta_1$ for excitons in direct band gap QDs has
been measured. The circular polarization degree for exciton emission
after optical orientation in indirect band gap QDs is determined by
the precession of the electron spin in the effective magnetic field
of the nuclear spin frozen fluctuations. The effect of this
hyperfine field of the frozen fluctuation of the nuclear spins on
the circular polarization degree is overcome by a longitudinal field
of 6 mT, resulting in $\rho_\text{c}=89\%$. The large
$\rho_\text{c}$ value shows that the exciton spin relaxation time
$T_1$ is much longer than the exciton radiative recombination time
$\tau_\text{R}$, which exceeds 30~$\mu$s.

\section*{Acknowledgements}
The authors are grateful to D. Dunker and J. Debus for their contribution in initial stages of this work. This
work was supported by the Deutsche Forschungsgemeinschaft via the
project No.~409810106 and in the frame of the International
Collaborative Research Center TRR 160 (Projects A1, B2 and B4), by
the Russian Foundation for Basic Research (Grant Nos. 19-02-00098
and 19-52-12001), by the Russian Federation government Grant No.
14.Z50.31.0021 (leading scientist M. Bayer), and by Act 211
Government of the Russian Federation (Contract No. 02.A03.21.0006).
Yu.G.K. and N.S.V. thank the Russian Science Foundation for support
of the experimental studies performed in the Ioffe Institute via
Grant No. 18-12-00352. P.K. acknowledges the support of his theory
studies by the national funding from the MEYS and the funding from
European Union's Horizon 2020 (2014-2020) research and innovation
framework programme under grant agreement No. 731473 and by project
EMPIR 17FUN06 SIQUST. This project received funding from the EMPIR
programme co-financed by the Participating States and from the
European Union Horizon 2020 research and innovation programme.

\end{document}